\documentclass[aps,prb,twocolumn,groupedaddress,showpacs,amsmath,amssymb,amsfonts]{revtex4}

\usepackage{latexsym,amssymb}
\usepackage{graphicx,color,pstricks}
\usepackage{epsfig,epsf,bm}

\definecolor{gray}{rgb}{0.7,0.7,0.7}

\begin{document}

\title{Interplay between Coulomb blockade and Josephson effect in a topological superconductor-quantum dot
device}

\author{Yu-Li Lee}
\email{yllee@cc.ncue.edu.tw}
\affiliation{Department of Physics, National Changhua University of Education, Changhua, Taiwan, R.O.C.}

\author{Yu-Wen Lee}
\email{ywlee@thu.edu.tw}
\affiliation{Department of Applied Physics, Tunghai University, Taichung, Taiwan, R.O.C.}

\date{\today}

\begin{abstract}
We study the behavior of a topological Josephson junction in which two topological superconductors are
coupled through a quantum dot. We focus on the case with the bulk superconducting gap being the largest
energy scale. Two parameter regimes are investigated: a weak tunneling between the dot and the
superconductors, with the dot near its charge degeneracy point, and a strong tunneling regime in which the
transmission between the dot and the superconductors is nearly perfect. We show that in the former
situation, the Andreev spectrum for each sector with fixed fermion parity consists of only two levels,
which gives rise to the nontrivial current-phase relation. Moreover, we study the Rabi oscillation between
the two levels and indicates that the corresponding frequency is a $4\pi$-periodic function of the phase
difference between the two superconductors, which is immune to the quasiparticle poisoning. In the latter
case, we find that the Coulomb charging energy enhances the effect of backscattering at the interfaces
between the dot and the superconductors. Both the temperature and the gate-voltage dependence of the
critical Josephson current are examined.
\end{abstract}


\maketitle

\section{Introduction}

Over the past few years, the Majorana bound states (MBSs) in topological superconductors (TSCs) have
attracted a lot of attention in the condensed matter research
community\cite{Alicea2,Leijnse,Beenakker,Beenakker2}. These exotic objects are basically the zero-energy
quasiparticles which are the equal-amplitude superposition of particles and holes in TSCs. One surprising
characteristic of these MBSs is that they obey non-Abelian braiding statistics\cite{Ivanov}. Furthermore,
a widely separated pair of MBSs forms a nonlocal fermionic state which is immune to local sources of
decoherence. Both features provide a potential for future applications in quantum computing and quantum
information processes\cite{Kitaev,Nayak,Alicea}. Up to now, there are several candidate systems that are
promised to realize these MBSs. Typically, s-wave superconductors are put in proximity to other materials
with strong spin-orbital coupling, in the presence or absence of external magnetic
fields\cite{Alicea2,Leijnse,Beenakker}. There have already been several recent experiments investigating
these kinds of systems\cite{Mourik,ADas,MTDeng,HOHChurchill}.

One way to reveal the existence of the MBSs is through the $4\pi$ (or fractional) Josephson effect in a
topological Josephson junction\cite{Kitaev,LFu,Lutchyn,KTLaw}, where the DC Josephson current is a
$4\pi$-periodic function of the superconducting (SC) phase difference between the two TSCs. This doubling
of the period is tightly related to these zero-energy MBSs. Roughly speaking, the existence of zero-energy
MBSs allow the coherent single-particle tunneling between TSCs. Since the electron carries half of the SC
phase, this results in the doubling of the period. However, in real experimental situations, there are
many subtle effects, such as disorder, multiple bands, quasi-particle poisoning, Coulomb charging energy
for SC islands etc., which have to be taken into
account\cite{Beenakker2,KTLaw,Sticlet,Pientka,Houzet,Heck}. Therefore, it is extremely important to
understand how this $4\pi$ Josephson effect survives these complications.

There are several recent attempts in this research direction. For example, by coupling the topological
Josephson junction to an external normal-metal probe, the dynamics of the fermion-parity switch through
the ejection of quasiparticles via the probe is analyzed in Ref. \onlinecite{Tara}. In Ref.
\onlinecite{Flensberg}, the effects of electromagnetic environments and an additional quantum dot (QD) on
the charge transport between the TSC and a normal-metal lead are considered. Interestingly, in the latter
case, it is claimed that when two TSCs are coupled via a QD, the resulting system is equivalent to a
resonant-level model at low energy from which a Coulomb oscillation in the conductance follows in both
the weak and the strong tunneling regime. It is not clear how the Josephson current, which is supposed to
exist in the absence of the QD, is affected by the presence of the QD.

\begin{figure}
\begin{center}
 \includegraphics[width=0.6\columnwidth]{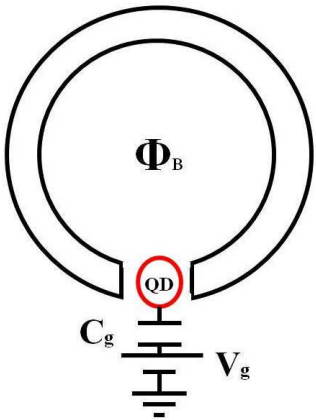}
 \caption{(Color online) A schematic setup of the TSC-QD-TSC junction. The TSC itself forms a ring
 threaded by a magnetic flux $\Phi_B$. The energy levels in the dot can be tuned by a capacitively coupled
 gate voltage $V_g$. The couplings between the QD and the TSC can be adjusted by additional gate
 voltages, which are not shown in the figure.}
 \label{ring}
\end{center}
\end{figure}

The interesting situation where the TSC is coupled to a metallic lead through a QD has already been
studied\cite{Leijnse2,AGoulb,MCheng,MLee}. There, the point is to analyze the competition between the
Andreev reflection and the Coulomb blockade/Kondo effect. In the present work, we would like to study the
effects of Coulomb blockade, introduced by a QD, on the DC Josephson current. The TSC itself forms a ring
with a gap, and the ring is threaded by a magnetic flux $\Phi_B$ which determines the SC phase difference
$\phi$ between the two ends of the TSC. The two ends of the TSC is coupled through a QD to form a
TSC-QD-TSC junction. A schematic setup is shown in Fig. \ref{ring}. We focus on the case where the bulk SC
gap is the largest energy scale. In the weak tunneling limit, we consider the case where the dot is near
the charge degeneracy point. This is because in this case the Andreev spectrum, arising from the
hybridization between the MBSs at the ends of the TSC and the charge degrees of freedom of the QD, shows a
simple structure, i.e. two Andreev levels for each sector with fixed fermion parity. This results in a
nontrivial current-phase relation with a $4\pi$ periodicity. This simple structure of the Andreev spectrum
and its $4\pi$ dependence on $\phi$ can be revealed as sharp peaks in the spectral function of the dot,
which could be measured by STM\cite{exp1} or microwave-optical experiments\cite{exp2}. Moreover, by
changing the flux through the ring abruptly, the Josephson current exhibits the behavior of Rabi
oscillation with the frequency being a $4\pi$-periodic function of $\phi$. This $\phi$ dependence of the
Rabi frequency is immune to quasiparticle poisoning. Both features, the $4\pi$ dependence of the peaks in
the spectral function of the dot and the Rabi frequency, can be viewed as the hallmark for the presence of
the MBSs. In the strong tunneling regime where the transmission between the QD and the SC electrodes is
nearly perfect, the current-phase relation is the usual one, i.e. $I=I_c\sin{(\phi/2)}$, provided that the
fermion parity is conserved. We calculate the temperature and the gate-voltage dependence of the critical
current $I_c$ by assuming that the temperature is much larger than the average level spacing in the dot.
Due to the Coulomb charging energy of the dot, the backscattering at the interfaces between the QD and the
SC electrodes is enhanced, which is revealed in the temperature dependence of $I_c$. Our main results are
summarized in Figs. \ref{sqdsi4} and \ref{sqdsi3}.

The rest of the paper is organized as follows. We present our analysis on the Andreev spectrum and the DC
Josephson current in the weak and the strong tunneling regimes in Sec. \ref{weak} and \ref{strong},
respectively. The last section is devoted to a conclusive discussion.

\section{The weak tunneling regime}
\label{weak}
\subsection{The model}

We first consider the weak tunneling regime, i.e. $G_{l/r}\ll e^2/h$, where $G_{l/r}$ denotes the
tunneling conductance for the left/right contact. At the energy scale much lower than the bulk SC gap
$\Delta_0$, the system can be described by the Hamiltonian $H=H_D+H_T$, where
\begin{equation}
 H_D=\! \sum_s\epsilon_sd^{\dagger}_sd_s+E_c(\hat{N}-N_g)^2 \ , \label{sqdsh1}
\end{equation}
is the Hamiltonian of the dot with $E_c=e^2/(2C)$ and
\begin{equation}
 H_T \! =\! \sum_s \! \left(\Gamma_{rs}e^{-i\chi_r/2}\gamma_rd_s-i\Gamma_{ls}e^{-i\chi_l/2}\gamma_ld_s+
 \mathrm{H.c.}\right) \! , \label{sqdsh11}
\end{equation}
describes the tunneling between the two ends of the TSC and the QD, with the real tunneling amplitudes
$\Gamma_{ls}$ and $\Gamma_{rs}$. We shall assume that $E_c,|\Gamma_{l(r)s}|\ll\Delta_0$. Here $\gamma_l$
and $\gamma_r$, obeying the algebra $\gamma_l^2=1=\gamma_r^2$ and $\{\gamma_l,\gamma_r\}=0$, denote the
two Majorana fermions at the ends of the left and the right SC electrodes, $\chi_l$ and $\chi_r$, obeying
the constraint $\chi_l-\chi_r=2e\Phi_B\equiv\phi$, are the SC phases of the left and the right SC
electrodes, $\epsilon_s$ denotes the single-particle energy of the dot with the complete set of good
quantum numbers $s$, the constant $N_g$ is proportional to the gate voltage $V_g$, the operators $d_s$ and
$d_s^{\dagger}$ obey the canonical anticommutation relations, and the Hermitian operator $\hat{N}$, having
integer eigenvalues, satisfies the commutation relations $[\hat{N},d_s]=-d_s$.

Let $|N\rangle$ be the eigenstate of the operator $\hat{N}$ with eigenvalues $N=0,\pm 1,\pm 2,\cdots$.
Following Ref. \onlinecite{LFu2}, for large charging energy, i.e. $E_c\gg|\Gamma_{l(r)s}|$, only two
charge states $|N_0\rangle$ and $|N_0+1\rangle$ give dominant contributions to the low-energy properties
near the degeneracy point $N_g=N_0+1/2$, where $N_0$ is some integer. To a good approximation, we can
truncate the Hilbert space keeping only these two states. Within this truncated space, one may introduce a
set of fermion operators $f$ and $f^{\dagger}$ defined by $f|N_0\rangle=0$ and
$f^{\dagger}|N_0\rangle=|N_0+1\rangle$. Moreover, they obey the canonical anticommutation relations. Then,
we have $\hat{N}-N_0=f^{\dagger}f$, and $H_D$ becomes
\begin{equation}
 H_D=\epsilon_df^{\dagger}f \ , \label{sqdsh12}
\end{equation}
up to a constant term, where $\epsilon_d$ accounts for the energy difference between the states
$|N_0+1\rangle$ and $|N_0\rangle$, and can be written as $\epsilon_d=\epsilon_0-2E_c(N_g-N_0-1/2)$. Here
the constant $\epsilon_0\equiv\langle N_0+1|\sum_s\epsilon_sd^{\dagger}_sd_s|N_0+1\rangle-\langle N_0|
\sum_s\epsilon_sd^{\dagger}_sd_s|N_0\rangle$ is independent of $N_g-N_0$. On the other hand, $H_T$ can be
written as
\begin{equation}
 H_T=t_re^{-i\chi_r/2}\gamma_rf-it_le^{-i\chi_l/2}\gamma_lf+\mathrm{H.c.} \ , \label{sqds13}
\end{equation}
where $t_{l(r)}\equiv\sum_s\Gamma_{l(r)s}\langle N_0|d_s|N_0+1\rangle$. Without loss of generality, we may
set $t_{l/r}>0$.

We see that near the degenerate point, the Hamiltonian $H=H_D+H_T$ is quadratic in the fermion operators
$f$ and $\gamma_{l/r}$, and thus the whole problem can be solved exactly. To proceed, we introduce another
set of fermion operators: $a=(\gamma_l+i\gamma_r)/2$ and $a^{\dagger}=(\gamma_l-i\gamma_r)/2$, which obey
the canonical anticommutation relations. Then, we may choose the set of orthonormalized states
$\{|0,0\rangle,|1,1\rangle,|0,1\rangle,|1,0\rangle\}$ as the basis of the truncated Hilbert space, where
the state $|m,n\rangle$ contains $m$ $f$-fermions and $n$ $a$-fermions. In terms of this choice of the
basis, the Hamiltonian $H$ can be written as
\begin{equation}
 H=\! \left[\begin{array}{cc}
 h_e & 0 \\
 0 & h_o
 \end{array}\right] , \label{sqdsh14}
\end{equation}
where
\begin{eqnarray}
 h_e &=& \! \left[\begin{array}{cc}
 0 & -i(\Delta_l+\Delta_r) \\
 & \\
 i(\Delta_l+\Delta_r)^* & \epsilon_d
 \end{array}\right] , \nonumber \\
 h_o &=& \! \left[\begin{array}{cc}
 0 & -i(\Delta_l-\Delta_r) \\
 & \\
 i(\Delta_l-\Delta_r)^* & \epsilon_d
 \end{array}\right] , \label{sqdsh15}
\end{eqnarray}
with $\Delta_{l/r}\equiv t_{l/r}e^{-i\chi_{l/r}/2}$. We see that $H$ is block diagonal due to the
fermion-parity symmetry of $H$. The Hermitian matrices $h_e$ and $h_o$ describe the dynamics of the states
in the subspaces with even and odd fermion parity, respectively.

\subsection{The Andreev spectrum and the DC Josephson current}

We first study the Andreev spectrum and calculate the resulting DC Josephson current. The energy spectrum
of $h_e$ is determined by its eigenvalues, which are given by
\begin{equation}
 E_{\pm}(\phi)=\frac{1}{2} \! \left(\epsilon_d\pm\sqrt{\epsilon_d^2+4|\Delta_l+\Delta_r|^2}\right) ,
 \label{sqdseq1}
\end{equation}
with the corresponding eigenstates denoted by $|\pm;\phi\rangle$, where
$|\Delta_l+\Delta_r|^2=t_l^2+t_r^2+2t_lt_r\cos{(\phi/2)}$. The Andreev spectrum will exhibit sharp peaks
in the spectral function of the dot, which is a $4\pi$-periodic function of $\phi$. This structure could
be measured by STM\cite{exp1} or microwave-optical experiments\cite{exp2}.

In terms of the formula $I_e=-2e\partial F_e/\partial\phi$, where $F_e$ is the free energy of the junction
with even fermion parity at finite temperature $T$, we get the DC Josephson current
\begin{equation}
 I_e(\phi)=I_{e0}(\phi)\tanh{\! \left[\frac{E_+(\phi)-E_-(\phi)}{2T}\right]} , \label{sqdseq13}
\end{equation}
where
\begin{equation}
 I_{e0}(\phi)=\frac{2et_lt_r\sin{(\phi/2)}}{\sqrt{\epsilon_d^2+4|\Delta_l+\Delta_r|^2}} \ ,
 \label{sqdseq14}
\end{equation}
is the DC Josephson current at $T=0$. We see that $I_e$ is a $4\pi$-periodic function of $\phi$, a
characteristic of the topological Josephson junction. In contrast with the short topological Josephson
junction in which the DC Josephson current is independent of $T$ when $T\ll\Delta_0$\cite{LFu}, $I_e$ is a
function of $T$ here.

\begin{figure}
\begin{center}
 \includegraphics[width=0.9\columnwidth]{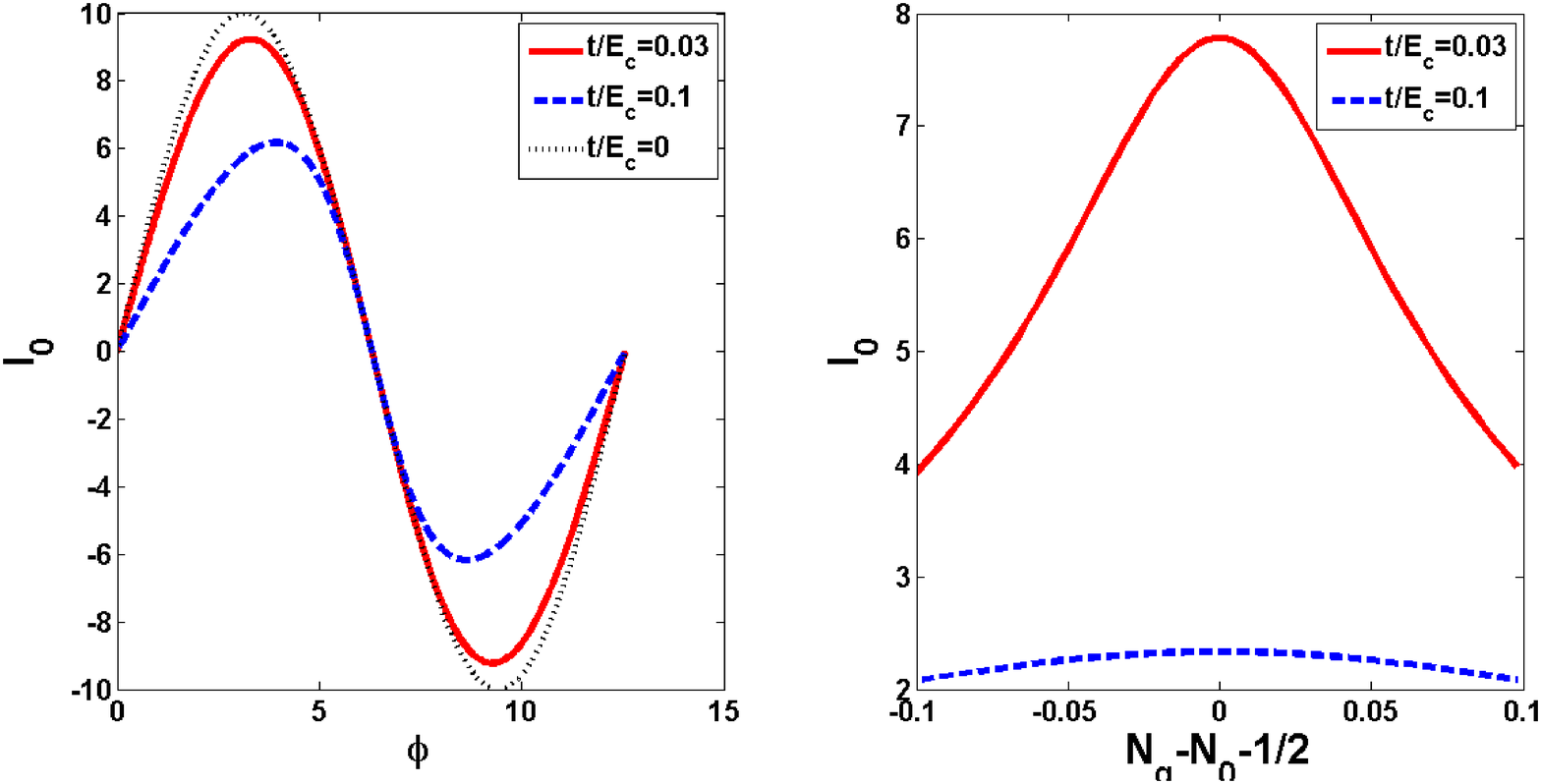}
 \caption{(Color online) The DC Josephson current $I_0$ at $T=0$ for the TSC-QD-TSC junction in the
 weak-tunneling regime with $N_g\approx N_0+1/2$. We set $t_l=t_r=t$, $\epsilon_0=0$, and measure $I_0$ in
 units of $et^2/E_c$. Only the results for the sector with even fermion parity are shown here. {\bf Left}:
 the current-phase relation at $N_g-N_0-1/2=\pm 0.1$ for $t/E_c=0.03$ (solid line) and $t/E_c=0.1$ (dashed
 line). We also plot $I_0$ in the limit $t/E_c\rightarrow 0$ to show the deviation from the one
 $I_0\propto\sin{(\phi/2)}$. {\bf Right}: $I_0$ versus $N_g-N_0-1/2$ at $\phi=0.3\pi$ for $t/E_c=0.03$
 (solid line) and $t/E_c=0.1$ (dashed line).}
 \label{sqdsi4}
\end{center}
\end{figure}

For the subspace with odd fermion parity, the Andreev spectrum is given by
\begin{equation}
 \tilde{E}_{\pm}(\phi)=\frac{1}{2} \! \left(\epsilon_d\pm\sqrt{\epsilon_d^2+4|\Delta_l-\Delta_r|^2}\right)
 , \label{sqdseq2}
\end{equation}
where $|\Delta_l-\Delta_r|^2=t_l^2+t_r^2-2t_lt_r\cos{(\phi/2)}$. Hence, the DC Josephson current is of the
form
\begin{equation}
 I_o(\phi)=I_{o0}(\phi)\tanh{\! \left[\frac{\tilde{E}_+(\phi)-\tilde{E}_-(\phi)}{2T}\right]} ,
 \label{sqdseq21}
\end{equation}
where
\begin{equation}
 I_{o0}(\phi)=-\frac{2et_lt_r\sin{(\phi/2)}}{\sqrt{\epsilon_d^2+4|\Delta_l-\Delta_r|^2}} \ ,
 \label{sqdseq22}
\end{equation}
is the DC Josephson current at $T=0$.

A few comments on the above results are in order. First of all, due to the presence of more than a single
Andreev level for each sector with fixed fermion parity, the current-phase relation at $T=0$ deviates from
the usual one proportional to $\sin{(\phi/2)}$. Next, for given $\phi$, $I_{e0}$ and $I_{o0}$ have
opposite signs, which is similar to the short topological Josephson junction\cite{LFu,KTLaw}. Finally, the
Andreev spectra for the sectors with even and odd fermion parity are related to each other through the
relation: $\tilde{E}_{\pm}(\phi\pm 2\pi)=E_{\pm}(\phi)$. Moreover, they cross each other at $\phi=\pi$,
leading to the fermion-parity anomaly as the short topological Josephson junction does. The behaviour of
$I_{e0}$ as a function of $\phi$ and $N_g$ is shown in Fig. \ref{sqdsi4}. Although our results hold only
when $N_g\approx N_0+1/2$, the DC Josephson current already shows the oscillatory behaviour due to the
charging energy.

\subsection{quenched dynamics}

Now we consider an abrupt change of the flux, i.e. $\phi=\phi_0$ and $\phi_0+\phi_1$ for $t<0$ and $t>0$,
respectively. This amounts to giving an initial state, say $|\Psi(0)\rangle=|+;\phi_0\rangle$, which is
not the eigenstates of $H(\phi_0+\phi_1)$. We would like to calculate the resulting Josephson current at
$t>0$.

The current operator $\hat{I}$ is defined as
\begin{equation}
 \hat{I}\equiv -2e\frac{\partial H}{\partial\phi} \ . \label{sqdsi1}
\end{equation}
With the choice of the initial state given above, the Josephson current at $T=0$ for the sector with even
fermion parity exhibits the behavior of Rabi oscillation:
\begin{equation}
 I_e(t)=I_0\cos{(\Omega_et+\alpha)} \ , \label{sqdsi11}
\end{equation}
where $I_0=e|(\Delta_l-\Delta_r)c_+^*c_-|$ is the current amplitude,
$\Omega_e=E_+-E_-=\sqrt{\epsilon_d^2+4|\Delta_l+\Delta_r|^2}$ is the Rabi frequency,
$\alpha\equiv\mbox{arg}[(\Delta_l-\Delta_r)c_+^*c_-]$, $c_{\pm}=\langle\pm ;\phi|+;\phi_0\rangle$, and
$\phi=\phi_0+\phi_1$. We notice that $I_0$, $\Omega_e$, and $\alpha$ are all periodic functions of $\phi$.
Especially, the period for $\Omega_e$ is $4\pi$, which is a signature of the zero-energy MBSs.

We would like to emphasize a few points on the above results. First of all, the Josephson current will
still exhibit the Rabi oscillation even if the change of $\Phi_B$ has a finite duration $\tau$. The
requirement is that $\tau\Omega_e\ll 1$. Next, by taking into account the dissipation that we have
ignored, the current will eventually decay into its equilibrium value given by Eq. (\ref{sqdseq14}).
However, the oscillatory behavior still exists before the equilibrium value is reached, and the frequency
$\Omega_e$ can be measured. Finally, in the presence of quasiparticle poisoning, the states with even and
odd fermion parity will both contribute to the Josephson current. Since the current operator $\hat{I}$
commutes with the fermion-parity operator, the contributions from the states with different fermion parity
will not mix, i.e. no interference appearing in the Josephson current. Therefore, the resulting current
still exhibits an oscillatory behavior with two Rabi frequencies $\Omega_e$ and $\Omega_o$, where
$\Omega_o=\tilde{E}_+-\tilde{E}_-=\sqrt{\epsilon_d^2+4|\Delta_l-\Delta_r|^2}$. Both are $4\pi$-periodic
functions of $\phi$. In this sense, the dependence of the Rabi frequencies on $\phi$ is immune to
quasiparticle poisoning.

\section{The strong tunneling regime}
\label{strong}
\subsection{The model}

Now we turn into the strong tunneling regime, i.e. $G_{l/r}\approx e^2/h$ or $\mathcal{T}_{l/r}\approx 1$,
where $\mathcal{T}_{l(r)}$ denotes the transmission coefficient of the left (right) contact. In this case,
the dot is connected to the SC electrodes through a narrow constriction. We assume that the width of the
constriction at its center allows only a single transverse state below the Fermi level. Since the
constriction is formed electrostatically, its boundary is smooth and do not scatter
electrons\cite{Glazmann}. Following the work of Furusaki and Matveev\cite{Matveev}, the whole system at
$\mathcal{T}_{l/r}=1$ amounts to a superconductor-normal-metal-superconductor (SNS) junction with a finite
charging energy. Hence, the Hamiltonian at $\mathcal{T}_{l/r}=1$ can be written as $H_0+H_c$, where at the
energy scale much lower than the bulk Fermi energy, $H_0$ can be written as\cite{MCheng2}
\begin{eqnarray}
 H_0 &=& \frac{v_F}{2} \! \int^{\infty}_{-\infty} \! dx \! \left[(\partial_x\Theta)^2
 +(\partial_x\Phi)^2\right] \nonumber \\
 & & +V_p \! \int^0_{-\infty} \! dx\cos{\! \left[\sqrt{4\pi}\Theta(x)+\chi_l\right]} \nonumber \\
 & & +V_p \! \int_L^{+\infty} \! dx\cos{\! \left[\sqrt{4\pi}\Theta(x)+\chi_r\right]} , \label{sqdsh2}
\end{eqnarray}
with $V_p=2\Delta_0\sin{(k_Fa_0)}/(\pi a_0)$, and
\begin{equation}
 H_c=E_c(\hat{N}-N_g)^2 \ , \label{sqdsh21}
\end{equation}
is the charging energy, with the exact form of $\hat{N}$ specified later. In the above, $v_F$ is the Fermi
velocity, $k_F$ is the Fermi momentum, $a_0$ is a short-distance cutoff, and
$\Delta_{l(r)}=\Delta_0e^{i\chi_{l(r)}}$ denotes the $p$-wave pairing amplitude in the left (right) SC
electrode with $\Delta_0>0$. Moreover, we take the center of the left contact and that of the right
contact to be located at $x=0$ and $x=L$, respectively. The bosonic fields $\Phi$ and $\Theta$, which obey
the commutation relation $[\Phi(x),\Theta(y)]=i\theta(y-x)$ with $\theta(x)$ being the Heaviside unit step
function, are related to the right mover $\psi_+$ and the left mover $\psi_-$ in the normal region through
the bosonization formula:
\begin{eqnarray*}
 \psi_{\pm}=\frac{1}{\sqrt{2\pi a_0}}\exp{\! \left[\pm i\sqrt{\pi}(\Phi\mp\Theta)\right]} .
\end{eqnarray*}
This model holds when $E_c\gg v_F/L$, where $v_F/L$ is the average level spacing in the QD. When
$\mathcal{T}_{l/r}$ are close to one, we have to add
a term $H_{bs}$ to $H_0+H_c$ to describe the weak backscattering at the contacts, where
\begin{eqnarray}
 & & \! \! \! \! H_{bs} =v_F \! \left[|r_l|\psi^{\dagger}_+(0)\psi_-(0)+|r_r|\psi^{\dagger}_+(L)\psi_-(L)+
     \mathrm{H.c.}\right] \nonumber \\
 & & \! \! \! \! =-\frac{v_F}{\pi a_0} \! \left\{|r_l|\sin{[\sqrt{4\pi}\Phi(0)]}+|r_r|
     \sin{[\sqrt{4\pi}\Phi(L)]}\right\} . \label{sqdsh22}
\end{eqnarray}
and $r_{l(r)}$ is reflection amplitude at the left (right) contact. We notice that $H_0$ [Eq.
(\ref{sqdsh2})] has been used in Ref. \onlinecite{MCheng2} to describe the topological SNS junction. In
our case, the Hamiltonian consists of an additional term -- the charging energy $H_c$. We shall see later
that the presence of $H_c$ will enhance the effects of $H_{bs}$, by turning it from an irrelevant
perturbation to a marginal one.

Without loss of generality, we may shift the value of $\Theta$ by
$\Theta(x)\rightarrow\Theta(x)-\chi_l/\sqrt{4\pi}$ such that $H_0$ can be written as
\begin{eqnarray}
 H_0 &=& \frac{v_F}{2} \! \int^{\infty}_{-\infty} \! dx \! \left[(\partial_x\Theta)^2
 +(\partial_x\Phi)^2\right] \nonumber \\
 & & +V_p \! \int^0_{-\infty} \! dx\cos{\! \left[\sqrt{4\pi}\Theta(x)\right]} \nonumber \\
 & & +V_p \! \int_L^{+\infty} \! dx\cos{\! \left[\sqrt{4\pi}\Theta(x)-\phi\right]} . \label{sqdsh23}
\end{eqnarray}
At the energy scale much smaller than the SC gap $\Delta_0$, the pairing terms suppress the fluctuations
of $\Theta(x)$ in the regions $x<0$ and $x>L$. Thus, for $E\ll\Delta_0$, one may integrate out the degrees
of freedom in these regions, and the resulting $H_0$ becomes
\begin{equation}
 H_0=\frac{v_F}{2} \! \int^L_0 \! dx \! \left[(\partial_x\Theta)^2+(\partial_x\Phi)^2\right] ,
 \label{sqdsh24}
\end{equation}
with the boundary conditions
\begin{equation}
 \Theta(0)=0 \ , ~~\Theta(L)=\frac{\phi}{\sqrt{4\pi}}-\sqrt{\pi}\hat{J} \ . \label{sqdsh25}
\end{equation}
In Eq. (\ref{sqdsh25}), $\hat{J}$ is a Hermitian operator with integer eigenvalues. The boundary
conditions (\ref{sqdsh25}), in fact, describe the perfect Andreev reflections at $x=0,L$, which is the
characteristic for the interface between the normal region and the TSC\cite{Fisher}.

Now we can write down the expression of the Hermitian operator $\hat{N}$, which is given by
\begin{equation}
 \hat{N}=\frac{1}{\sqrt{\pi}} \! \int^L_0 \! dx\partial_x\Phi=\frac{\Phi(L)-\Phi(0)}{\sqrt{\pi}} \ .
 \label{sqdsh26}
\end{equation}
Hence, $H_c$ becomes
\begin{equation}
 H_c=\frac{E_c}{\pi}[\Phi(L)-\Phi(0)-\sqrt{\pi}N_g]^2 \ . \label{sqdsh27}
\end{equation}
Our working Hamiltonian is $H=H_0+H_c+H_{bs}$ with $H_0$, $H_c$, and $H_{bs}$ given by Eqs.
(\ref{sqdsh24}), (\ref{sqdsh27}), and (\ref{sqdsh22}), respectively. Moreover, we shall assume that
$|r_{l/r}|\ll 1$.

The mode expansions of $\Phi(x)$ and $\Theta(x)$ consistent with the boundary conditions (\ref{sqdsh25})
are of the forms
\begin{eqnarray}
 \Phi(x) &=& -\frac{\hat{q}}{\sqrt{\pi}}+\bar{\Phi}(x) \ , \nonumber \\
 \Theta(x) &=& \! \left(\frac{\phi}{\sqrt{4\pi}}-\sqrt{\pi}\hat{J}\right) \! \frac{x}{L}+\bar{\Theta}(x)
 \ , \label{bmod0}
\end{eqnarray}
where the Hermitian operator $\hat{q}$ is conjugate to $\hat{J}$, i.e. $[\hat{q},\hat{J}]=i$, and
\begin{eqnarray}
 \bar{\Phi}(x) &=& \! \sum_{n=1}^{+\infty}\frac{i}{\sqrt{q_nL}}\cos{(q_nx)} \! \left(a_n-a_n^{\dagger}
 \right) , \nonumber \\
 \bar{\Theta}(x) &=& \! \sum_{n=1}^{+\infty}\frac{1}{\sqrt{q_nL}}\sin{(q_nx)}\! \left(a_n+a_n^{\dagger}
 \right) , \label{bmod1}
\end{eqnarray}
with $q_n=n\pi/L$. In the above, $a_n$ and $a_n^{\dagger}$ obey the canonical commutation relations.
Moreover, both commute with $\hat{q}$ and $\hat{J}$. Inserting Eqs. (\ref{bmod0}) and (\ref{bmod1}) into
Eq. (\ref{sqdsh24}) gives
\begin{equation}
 H_0=\! \sum_{n=1}^{+\infty}v_Fq_na_n^{\dagger}a_n+\frac{\pi v_F}{2L} \! \left(\hat{J}-\frac{\phi}{2\pi}
 \right)^2 . \label{sqdsh3}
\end{equation}
The second term in Eq. (\ref{sqdsh3}) describes the dynamics of the zero modes, while the first term
describes the excitations in the QD.

\subsection{The effective action of the zero modes}

From Eq. (\ref{sqdsh3}), we notice that $\phi$ appears only in the zero modes. Thus, to study the
Josephson effect, we would like to integrate out the $\bar{\Phi}$ field to get an effective Hamiltonian
for the zero modes. We do this by two steps. First of all, we notice that $H_c$ and $H_{bs}$ depend only
on $\Phi(0)$ and $\Phi(L)$. This observation suggests us to integrate out $\bar{\Phi}(x)$ with $x\neq 0,L$
to get an effective theory for $\phi_l=\bar{\Phi}(0)$ and $\phi_r=\bar{\Phi}(L)$.

To achieve this goal, we turn into the path-integral formulation. Since the action is Gaussian in
$\bar{\Phi}(x)$ with $x\neq 0,L$, the integral can be done exactly and the resulting effective action for
$\phi_l$ and $\phi_r$ is given by
\begin{eqnarray}
 S_{eff} &=& \frac{1}{2}\! \int^{\beta}_0 \! d\tau_1d\tau_2\phi_I(\tau_1)\Delta_I^{-1}(\tau_1-\tau_2)
 \phi_I(\tau_2) \nonumber \\
 & & +\frac{1}{2}\! \int^{\beta}_0 \! d\tau_1d\tau_2\phi_C(\tau_1)\Delta_C^{-1}(\tau_1-\tau_2)
 \phi_C(\tau_2) \nonumber \\
 & & -\lambda_l \! \int^{\beta}_0 \! d\tau\sin{[\sqrt{2\pi}(\phi_I-\phi_C)-2q]} \nonumber \\
 & & -\lambda_r \! \int^{\beta}_0 \! d\tau\sin{[\sqrt{2\pi}(\phi_I+\phi_C)-2q]} \nonumber \\
 & & +\frac{2E_c}{\pi} \! \int^{\beta}_0 \! d\tau \! \left(\phi_C-\sqrt{\frac{\pi}{2}}N_g\right)^2 ,
 \label{sqdss13}
\end{eqnarray}
where $\phi_l=(\phi_I-\phi_C)/\sqrt{2}$, $\phi_r=(\phi_I+\phi_C)/\sqrt{2}$,
$\lambda_{l/r}=v_F|r_{l/r}|/(\pi a_0)$, and
\begin{eqnarray*}
 \Delta_I(\tau) &=& D(\tau;0,0)+D(\tau;0,L) \ , \\
 \Delta_C(\tau) &=& D(\tau;0,0)-D(\tau;0,L) \ .
\end{eqnarray*}
In the above,
\begin{eqnarray*}
 D(\tau;x_1,x_2)\equiv\langle\mathcal{T}_{\tau}\{\bar{\Phi}(\tau_1,x_1)\bar{\Phi}(\tau_2,x_2)\}\rangle_0
 \ ,
\end{eqnarray*}
is the free propagator of $\bar{\Phi}$, where$\tau\equiv\tau_1-\tau_2$ and $\mathcal{T}_{\tau}$ denotes
time ordering along the imaginary-time axis. For the temperature $T\gg v_F/L$, we have
\begin{eqnarray*}
 \tilde{\Delta}_I(i\omega_l)\approx \frac{1}{|\omega_l|} \ , ~~
 \tilde{\Delta}_C(i\omega_l)\approx \frac{1}{|\omega_l|} \ ,
\end{eqnarray*}
where $\omega_l=2l\pi T$ with $l$ being integer and $\tilde{A}(i\omega_l)$ is the Fourier transform of
$A(\tau)$.

At the energy scale much smaller than $E_c$, the charging energy $H_c$ suppresses the fluctuations of
$\phi_C$, by setting its average value to be $\langle\phi_C\rangle=\sqrt{\pi/2}N_g$. Thus, at energy
scales much smaller than $E_c$, one may integrate out the fluctuations of $\phi_C$ by replacing the
scattering terms in $S_{eff}$ [Eq. (\ref{sqdss13})] with its values averaged over the Gaussian
fluctuations of $\phi_C$, yielding the effective action for $\phi_I$
\begin{eqnarray}
 \tilde{S} &=& \frac{1}{2} \! \int^{\beta}_0 \! d\tau_1d\tau_2\phi_I(\tau_1)\Delta_I^{-1}(\tau_1-\tau_2)
 \phi_I(\tau_2) \nonumber \\
 & & +\! \int^{\beta}_0 \! d\tau \! \left[\tilde{\lambda}e^{i(\sqrt{2\pi}\phi_I-2q)}+\mathrm{H.c.}\right]
 , \label{sqdss14}
\end{eqnarray}
where $\gamma=0.5772156649\cdots$ is the Euler's constant and
\begin{eqnarray*}
 \tilde{\lambda}=\frac{2ie^{\gamma}E_c}{\pi^2} \! \left(|r_l|e^{-i\pi N_g}+|r_r|e^{i\pi N_g}\right) .
\end{eqnarray*}
This is, in fact, a perturbative expansion in $\lambda_{l/r}$, which is justified because $|r_{l/r}|\ll 1$
and the fluctuations of $\phi_C$ around its expectation value are gapped.

Now we are in a position to integrate out $\phi_I$. We notice that
$\tilde{\Delta}^{-1}_I(i\omega_l)=|\omega_l|$, which implies that the scaling dimension of the
$\tilde{\lambda}$ term is $1$. That is, it is a marginal perturbation. (In the absence of the charging
energy $H_c$, the scaling dimension of $H_{bs}$ is $2$ so that it is an irrelevant
perturbation\cite{Fisher}. In this sense, the large charging energy enhances the effects of $H_{bs}$ on
low-energy physics by freezing parts of the charge fluctuations.) Since $|r_{l/r}|\ll 1$, we may integrate
out $\phi_I$ by the perturbative expansion in $\tilde{\lambda}$, yielding the effective action for the
zero modes:
\begin{equation}
 \mathcal{I}=\mathcal{I}_0+\sum_{n=1}^{+\infty}\delta\mathcal{I}_n \ , \label{sqdss2}
\end{equation}
where $\delta\mathcal{I}_n=O(|\tilde{\lambda}|^{2n})$ and
\begin{equation}
 \mathcal{I}_0=\! \int^{\beta}_0 \! d\tau \! \left[J(-i\partial_{\tau}q)+\frac{\pi v_F}{2L} \! \left(
 J-\frac{\phi}{2\pi}\right)^2\right] . \label{sqdss21}
\end{equation}
The partition function of the TSC-QD-TSC junction is then given by
\begin{equation}
 Z=\! \int \! D[q]\sum_{J(\tau)=-\infty}^{+\infty}e^{-\mathcal{I}} \ . \label{sqdss22}
\end{equation}
Since the eigenvalues of $\hat{J}$ take integer values, $q$ must be an angular variable. Without loss of
generality, we take $0\leq q<2\pi$ and the periodic boundary condition, i.e. $q(\beta)=q(0)$ in Eq.
(\ref{sqdss22}).

It is convenient to calculate the free energy in terms of a dual representation of Eq. (\ref{sqdss22}). To
achieve this goal, we first perform the summation over $J$. By using the Hubbard-Stratonovich
transformation, we find that
\begin{eqnarray*}
 \sum_{J(\tau)=-\infty}^{+\infty}e^{-\mathcal{I}_0}=\! \int \! D[\rho]
 e^{-\frac{L}{2\pi v_F}\rho^2+i\frac{\phi}{2\pi}\rho}Q[\rho,q] \ ,
\end{eqnarray*}
where
\begin{eqnarray*}
 & & Q[\rho,q]=\! \sum_{J(\tau)=-\infty}^{+\infty}
     \exp{\! \left[i \! \int^{\beta}_0 \! d\tau J(\partial_{\tau}q-\rho)\right]} \\
 & & =\! \! \sum_{J_j=-\infty}^{+\infty} \!
     \exp{\! \left[i\lim_{N\rightarrow +\infty}\! \sum_{j=0}^{N-1}J_j(q_{j+1}-b_{j+1}-q_j+b_j)\right]}
     \\
 & & =\lim_{N\rightarrow +\infty} \! \prod_{j=0}^{N-1}\! \sum_{J_j=-\infty}^{+\infty} \!
     e^{iJ_j(q_{j+1}-b_{j+1}-q_j+b_j)} \ ,
\end{eqnarray*}
with $A_j=A(\tau_j)$, $\Delta\tau=\beta/N$, and $\tau_j=j\Delta\tau$. Moreover, we have written
$\rho=\partial_{\tau}b$. By using the Poisson summation formula
\begin{eqnarray*}
 \sum_{n=-\infty}^{\infty}\delta (x-2\pi n)=\frac{1}{2\pi}\sum_{m=-\infty}^{\infty}e^{imx} \ .
\end{eqnarray*}
we get
\begin{eqnarray*}
 Q[\rho,q] \! =\! \! \lim_{N\rightarrow +\infty}\! \prod_{j=0}^{N-1}\! \sum_{l=-\infty}^{+\infty} \! \! \!
 \delta \! \left(\frac{q_{j+1}-b_{j+1}-q_j+b_j}{2\pi}-l\right) \! .
\end{eqnarray*}
We see that $Q[\rho,q]$ provides each time slice a $\delta$-function which imposes the constraints
$b_{j+1}-b_j=q_{j+1}-q_j-2\pi l_j$. The integration over $b$ in the presence of these local constraints
leads to a global one $b(\beta)-b(0)=2m\pi$ where $m=-\sum_{j=1}^{N-1}l_j$ is an integer. Since
$\rho(\beta)=\rho(0)$, this is possible only when $b(\tau)=q(\tau)+2m\pi\tau/\beta$. Because $b$ appears
in the action only in the guise of $\partial_{\tau}b$, we may rewrite $Q[\rho,q]$ as
\begin{eqnarray*}
 Q[\rho,q]=\sum_{m=-\infty}^{+\infty}\delta \! \left[b(\tau)-q(\tau)-\frac{2m\pi\tau}{\beta}\right] .
\end{eqnarray*}

Now we are in a position to integrate out $b$, and the partition function $Z$ becomes
\begin{equation}
 Z=\! \sum_{m=-\infty}^{+\infty} \! \int_{q(\beta)-q(0)=2m\pi} \! D[q]e^{-\tilde{I}} \ , \label{sqdss23}
\end{equation}
where
\begin{equation}
 \tilde{I}=\! \int^{\beta}_0 \! d\tau \! \left[\frac{L}{2\pi v_F}(\partial_{\tau}q)^2-i\frac{\phi}
 {2\pi}\partial_{\tau}q\right] \! +\sum_{n=1}^{+\infty}\delta\mathcal{I}_n \ . \label{sqdss24}
\end{equation}
The effects of the term $2m\pi\tau/\beta$ have been taken into account by changing the boundary condition
for $q(\tau)$ from the periodic one to $q(\beta)-q(0)=2m\pi$. Equation (\ref{sqdss23}) is the desired dual
representation of Eq. (\ref{sqdss22}). We shall use both in the following.

\subsection{The DC Josephson current}

Now we are able to compute the DC Josephson current. When $\tilde{\lambda}=0$ or $\mathcal{T}_{l/r}=1$,
the action $\tilde{I}$ is nothing but the one for a particle, of mass $L/(\pi v_F)$ and unit charge,
moving on a unit circle thread by a flux $\phi$, provided that we treat $q$ as the azimuthal angle. In
this sense, the integer $m$ is the winding number. Hence the partition function at $\mathcal{T}_{l/r}=1$
can be written as
\begin{equation}
 Z_0(\phi)=\sum_{J=-\infty}^{+\infty}e^{-\beta\epsilon_J} \ , \label{sqdsz1}
\end{equation}
leading to the free energy $F_0=-T\ln{Z_0}$, where
$\epsilon_J=\frac{\pi v_F}{2L}[J-\phi/(2\pi)]^2$ is the energy of the zero mode with the quantum number
$J$. The resulting DC Josephson current is of the form
\begin{equation}
 I^{(0)}(\phi)=\frac{ev_F}{L}\sum_{J=-\infty}^{+\infty} \! \left(J-\frac{\phi}{2\pi}\right) \! P_J \ ,
 \label{sqdsi2}
\end{equation}
where $P_J=\frac{1}{Z_0}e^{-\beta\epsilon_J}$ is the probability of the zero mode at the eigenstate
$|J\rangle$. We notice that the whole spectrum is $2\pi$-periodic in $\phi$, and thus $I^{(0)}$ is a
$2\pi$-periodic function of $\phi$. Moreover, $I^{(0)}(-\phi)=-I^{(0)}(\phi)$, which implies that
$I^{(0)}(0)=0$. The eigenvalue of $\hat{J}$ is the difference between the numbers of the right- and the
left-movers. Hence, even and odd values of $J$ correspond to different fermion parities. By taking into
account the conservation of fermion parity, the allowed values of $J$ become either odd or even integers.
Consequently, $I^{(0)}$ turns into a $4\pi$-periodic function of $\phi$, a characteristic of topological
Josephson junctions.

To proceed, it is more convenient to use the dual representation of $Z_0$ [Eq. (\ref{sqdss23})]. By
writing $q(\tau)=\bar{q}(\tau)+2m\pi\tau/\beta$ with $\bar{q}(\beta)=\bar{q}(0)$ and then performing the
integration over $\bar{q}$, we obtain $Z_0$
\begin{equation}
 Z_0(\phi)=\sqrt{\frac{2L}{v_F\beta}} \! \sum_{m=-\infty}^{+\infty}
 e^{im\phi-\frac{L}{2\pi v_F\beta}(2m\pi)^2} \ . \label{sqdsz11}
\end{equation}
Since $T\gg v_F/L$, the dominant contributions arise from the terms with $m=0$ and $m=\pm 1$. Hence, we
may approximate $Z_0$ as
\begin{equation}
 Z_0(\phi)\approx\sqrt{\frac{2L}{v_F\beta}} \! \left(1+2e^{-\frac{2\pi L}{v_F\beta}}\cos{\phi}\right) .
 \label{sqdsz12}
\end{equation}
Consequently, the DC Josephson current is given by
\begin{equation}
 I^{(0)}(\phi)\approx 4eTe^{-\frac{2\pi L}{v_F\beta}}\sin{\phi} \ , \label{sqdsi21}
\end{equation}
which holds when $v_F/L\ll T\ll E_c$.

In the above calculation, we do not take into account the conservation of fermion parity so that $I^{(0)}$
is a $2\pi$-periodic function of $\phi$. The constraint from the fermion-parity conservation can be
included by assuming that $J$ must be even (or odd) integers. With the similar procedure, we find that
\begin{equation}
 I^{(0)}_e(\phi)\approx 2eTe^{-\frac{\pi L}{2v_F\beta}}\sin{(\phi/2)}=-I^{(0)}_o(\phi) \ . \label{sqdsi20}
\end{equation}
We see that $I^{(0)}_e$ and $I^{(0)}_o$ have opposite signs at given $T$ and $\phi$.

When $\tilde{\lambda}\neq 0$, small reflection amplitudes at the contacts will modify the DC Josephson
current. We shall calculate the corrections to $\mathcal{I}^{(0)}$ by the perturbative expansion in
$|\tilde{\lambda}|^2$. The leading contribution arises from $\delta\mathcal{I}_1$ which takes the form
\begin{eqnarray*}
 \delta\mathcal{I}_1 \! =-|\tilde{\lambda}|^2 \! \! \int^{\beta}_0 \! d\tau_1d\tau_2V(\tau_1-\tau_2)
 \cos{\{2[q(\tau_1)-q(\tau_2)]\}} ,
\end{eqnarray*}
where
\begin{eqnarray*}
 V(\tau) &=& \! \left\langle e^{i\sqrt{2\pi}\phi_I(\tau)}e^{-i\sqrt{2\pi}\phi_I(0)}\right\rangle \\
 &=& \exp{\! \left\{2\pi[\Delta_I(\tau)-\Delta_I(0)]\right\}} .
\end{eqnarray*}
Using the Baker-Hausdorff formula, one may verify that
\begin{eqnarray*}
 [\hat{J},e^{i\alpha\hat{q}}]=\alpha e^{i\alpha\hat{q}} \ .
\end{eqnarray*}
This equation indicates that the operator $e^{i\alpha\hat{q}}$ shifts the eigenvalue of $\hat{J}$ by
$\alpha$. Thus, the effects of $\delta\mathcal{I}_1$ is to shift the eigenvalue of $\hat{J}$ by $\pm 2$.
In general, the terms in $\delta\mathcal{I}_n$ will shift the eigenvalue of $\hat{J}$ by $\pm 2n$. As a
result, all the terms $\delta\mathcal{I}_n$ do not change the fermion parity.

In general, a perturbative expansion of the free energy $F$ can be written as
\begin{eqnarray*}
 F=F_0+\sum_{n=1}^{+\infty}F_n \ ,
\end{eqnarray*}
where $F_n=O(|\tilde{\lambda}|^{2n})$. In particular, we have
\begin{eqnarray*}
 F_1=T\langle\delta\mathcal{I}_1\rangle_0=-T|\tilde{\lambda}|^2 \! \! \int^{\beta}_0 \!  d\tau_1d\tau_2
 V(\tau)C(\tau;\phi) \ ,
\end{eqnarray*}
where $\langle\cdots\rangle_0$ denotes the average with respect to $Z_0$, $\tau=\tau_1-\tau_2$, and
\begin{eqnarray*}
 C(\tau;\phi)=\langle\cos{\{2[\hat{q}(\tau_1)-\hat{q}(\tau_2)]\}}\rangle_0 \ .
\end{eqnarray*}
When $T\gg v_F/L$, $C(\tau;\phi)$ can be approximated as
\begin{eqnarray*}
 C(\tau;\phi) &\approx& \! \left[1-4e^{-\frac{2\pi L}{v_F\beta}}\cos{\phi}
 \sin^2{\! \left(\frac{2\pi \tau}{\beta}\right)}\right] \\
 & & \times\exp{\! \left[-\frac{2\pi v_F}{\beta L}\tau(\beta-\tau)\right]} ,
\end{eqnarray*}
for $0\leq\tau\leq\beta$.

\begin{figure}
\begin{center}
 \includegraphics[width=0.9\columnwidth]{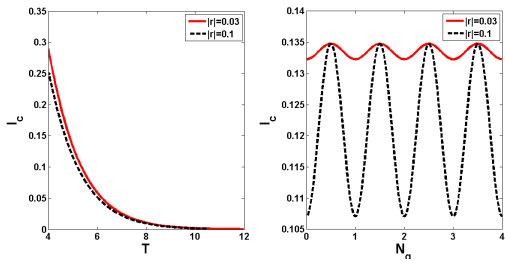}
 \caption{(Color online) The critical current $I_c$ [Eq. (\ref{sqdsi23})] for a TSC-QD-TSC junction in
 the strong-tunneling regime. We set $|r_l|=|r_r|=|r|$ and measure $I_c$ in units of $ev_F/(2\pi L)$.
 {\bf Left}: $I_c$ versus the temperature $T$ (in units of $v_F/(2\pi L)$) at $N_g=0.2$ for $|r|=0.03$
 (solid line) and $|r|=0.1$ (dashed line). {\bf Right}: $I_c$ versus $N_g$ at $T=5v_F/(2\pi L)$ for
 $|r|=0.03$ (solid line) and $|r|=0.1$ (dashed line).}
 \label{sqdsi3}
\end{center}
\end{figure}

Collecting the above results, $F_1$ is given by
\begin{eqnarray*}
 F_1 &\approx& -2T|\tilde{\lambda}|^2 \! \int^{\beta}_0 \! d\tau\tau V(\tau)
 \exp{\! \left[-\frac{2\pi v_F}{\beta L}\tau(\beta-\tau)\right]} \\
 & & \times \! \left[1-4e^{-\frac{2\pi L}{v_F\beta}}\cos{\phi}
 \sin^2{\! \left(\frac{2\pi \tau}{\beta}\right)}\right] ,
\end{eqnarray*}
when $v_F/L\ll T\ll E_c$, which leads to the leading correction to the DC Josephson current
\begin{eqnarray*}
 I^{(1)}=-\frac{16\pi^2eT|\tilde{\lambda}|^2}{E_c^2}e^{-\frac{2\pi L}{v_F\beta}}\sin{\phi} \ .
\end{eqnarray*}
As a result, we find that for $v_F/L\ll T\ll E_c$, the DC Josephson current is of the form
\begin{equation}
 I=I_{2\pi}\sin{\phi} \ , \label{sqdsi22}
\end{equation}
where the critical current $I_{2\pi}$ is given by
\begin{equation}
 I_{2\pi}=4eTe^{-\frac{2\pi L}{v_F\beta}} \! \left(1-\frac{4\pi^2|\tilde{\lambda}|^2}{E_c^2}\right) ,
 \label{sqdsi23}
\end{equation}
to the leading order in $|\tilde{\lambda}|^2$. We see that the small reflection amplitudes at the contacts
do not change the current-phase relation. They manifests themselves in the critical current in two ways.
(i) First of all, the small reflection amplitudes at the contacts diminish the critical current. (ii)
Next, the current acquires the dependence on the gate voltage through the reflection amplitudes because
\begin{eqnarray*}
 \frac{4\pi^2|\tilde{\lambda}|^2}{E_c^2}=\! \left(\frac{4e^{\gamma}}{\pi}\right)^{\! \! 2} \! \! \left[
 |r_l|^2+|r_r|^2+2|r_l||r_r|\cos{(2\pi N_g)}\right] .
\end{eqnarray*}
Hence, at given temperature, the critical current reaches the maximal values for half-integer $N_g$ and
the minimal values for integer $N_g$. The behaviours of $I_{2\pi}$ versus $T$ and $N_g$ are shown in Figs.
\ref{sqdsi3}. By comparing with the results in the weak-tunneling regime, we conclude that the
current-phase relation for the TSC-QD-TSC junction can be changed by varying the values of
$\mathcal{T}_{l/r}$, which can be achieved by adjusting the applied gate voltages which form the
constrictions.

We notice that $I^{(1)}$ has the same temperature dependence as $I^{(0)}$. This reflects the fact that
$H_{bs}$ is a marginal perturbation, which follows from the large charging energy. If there were no
charging energy, $H_{bs}$ would be an irrelevant perturbation such that an additional factor $T$ would
appear in $I^{(1)}$. Consequently, the effects of $H_{bs}$ would be suppressed at low
temperatures\cite{Fisher}. Our analysis indicates that the effects of reflections at the contacts are
enhanced in the presence of a large charging energy.

In the above calculation, we do not take into account the conservation of fermion parity. For fixed
fermion parity, the only change in $F_1$ lies in the function $C(\tau;\phi)$, and we obtain
\begin{equation}
 I_e=I_{4\pi}\sin{(\phi/2)}=-I_o \ , \label{sqdsi24}
\end{equation}
where
\begin{equation}
 I_{4\pi}=2eTe^{-\frac{\pi L}{2v_F\beta}} \! \left(1-\frac{4\pi^2|\tilde{\lambda}|^2}{E_c^2}\right) ,
 \label{sqdsi25}
\end{equation}
is the critical current with fixed fermion parity. We see that $I_e$ and $I_o$ have opposite signs at
given $T$, $\phi$, and $N_g$. Moreover, the ratio $I_{4\pi}/I_{2\pi}$ depends on the temperature $T$ in a
nontrivial way, but is independent of the gate voltage.

\section{Conclusions and discussions}

We study the DC Josephson effect in a TSC-QD-TSC junction. In the weak-tunneling regime, the Andreev
spectrum becomes simple near the degenerate point of the QD. This simple structure should appear as sharp
peaks in the spectral function of the dot, which could be observed by established experimental techniques.
We also consider the abrupt change of the flux, which results in a Rabi oscillation of the Josephson
current with frequencies being $4\pi$-periodic functions of $\phi$. We indicate that this feature is
immune to the quasiparticle poisoning. Far away from the degenerate point, the proliferation of Andreev
levels will complicate the behavior of the Josephson current. Therefore, we do not consider such a
situation. We notice that the use of the Rabi oscillation to extract the signature of the MBSs has been
proposed in a different setup\cite{XHu}. In that case, the QD is side-coupled to one end of the TSC in a
Josephson junction, and it is the electron occupation of the dot which exhibits the Rabi oscillation due
to the DC Josephson current.

In the strong-tunneling regime, we model the dot and the contacts as a one-dimensional electron liquid,
following the work on the charge transport through an open QD\cite{Matveev}. Using the bosonization, we
may calculate the free energy and the DC Josephson current in a perturbative expansion in the reflection
amplitudes. The resulting current-phase relation is of the form $I_c\sin{(\phi/2)}$, identical to the
short Josephson junction. The difference lies at the critical current $I_c$, which acquires a nontrivial
dependence on the temperature and the gate voltage. In the absence of the backscattering terms, the
critical current we obtained is identical to the one for the long topological SNS junction in the
temperature range $T\gg v_F/L$\cite{Beenakker3}.

In the absence of the charging energy $H_c$, the low-energy effective Hamiltonian is identical to the one
for the long topological SNS junction\cite{MCheng2,Crepin}. The role of the charging energy is to enhance
the effects of the backscattering at the contacts on low-energy properties. This has been shown in the
study of the charge transport through the open QD\cite{Matveev}. There, the backscattering terms become
relevant perturbations and a non-perturbative method is warranted. In the present case, thanks to the
superconductivity in the bulk, the backscattering terms turn into a marginal perturbation, and a
perturbative expansion works.

In the weak tunneling regime, we only consider the case near the degenerate point of the QD so that only 
two levels of the dot are involved. It is interesting to study how the Andreev spectrum and the resulting 
current-phase relation evolve upon moving far away from the degenerate point. On the other hand, our 
results in the strong tunneling regime hold only for $T\gg v_F/L$. Thus, the DC Josephson effect at the 
temperature range $T\approx v_F/L$ requires further study. Finally, as our analysis suggests, the 
current-phase relation of the TSC-QD-TSC junction depends on the transmission coefficients 
$\mathcal{T}_{l/r}$ of the contacts. Since the control of the critical current is important for the basic
elements in large-scale integrated SC circuits\cite{Golubov}, the dependence of the current-phase relation 
on $\mathcal{T}_{l/r}$ is deserved to investigate.

\acknowledgments

The work of Y.-W. Lee is supported by the National Science Council of Taiwan under Grant No. NSC
102-2112-M-029-002-MY3.


\end{document}